\documentclass[twocolumn,showpacs,preprintnumbers,amsmath,amssymb]{revtex4}

\usepackage{graphicx}
\usepackage{dcolumn}
\usepackage{bm}
\usepackage{color}

\begin{document}

\title{
Exact diagonalization and cluster mean-field study of triangular-lattice XXZ antiferromagnets near saturation}

\author{Daisuke Yamamoto$^1$, Hiroshi Ueda$^2$, Ippei Danshita$^{3}$, Giacomo Marmorini$^{3,4}$, Tsutomu Momoi$^{5,6}$, and Tokuro Shimokawa$^{7}$}
\affiliation{$^1$Department of Physics and Mathematics, Aoyama-Gakuin University, Sagamihara, Kanagawa 252-5258, Japan}
\affiliation{$^2$RIKEN Advanced Institute for Computational Science (AICS), Kobe, Hyogo 650-0047, Japan}
\affiliation{$^3$Yukawa Institute for Theoretical Physics, Kyoto University, Kyoto 606-8502, Japan}
\affiliation{$^4$Research and Education Center for Natural Sciences, Keio University, Kanagawa 223-8521, Japan}
\affiliation{$^5$Center for Emergent Matter Science (CEMS), RIKEN, Wako, Saitama 351-0198, Japan}
\affiliation{$^6$Condensed Matter Theory Laboratory, RIKEN, Wako, Saitama 351-0198, Japan}
\affiliation{$^7${Okinawa Institute of Science and Technology Graduate University, Onna, Okinawa, 904-0495, Japan}.}
\date{\today}

\begin{abstract}
Quantum magnetic phases near the magnetic saturation of triangular-lattice antiferromagnets with XXZ anisotropy have been attracting renewed interest since it has been suggested that a nontrivial coplanar phase, called the $\pi$-coplanar or $\Psi$ phase, could be stabilized by quantum effects in a certain range of anisotropy parameter $J/J_z$ {besides} the well-known 0-coplanar (known also as $V$) and umbrella phases. {Recently, Sellmann $et$ $al$. [Phys. Rev. B {\bf 91}, 081104(R) (2015)] claimed that the $\pi$-coplanar phase is absent for $S=1/2$ from an exact-diagonalization analysis in the sector of the Hilbert space with only three down-spins (three magnons). We first reconsider and improve this analysis by taking into account several low-lying eigenvalues and the associated eigenstates as a function of $J/J_z$ and by sensibly increasing the system sizes (up to 1296 spins). A careful identification} analysis shows that the lowest eigenstate is a chirally antisymmetric combination of finite-size umbrella states for {$J/J_z\gtrsim 2.218$} while it corresponds to a coplanar phase for {$J/J_z\lesssim 2.218$}. {However, we demonstrate that the distinction between 0-coplanar and $\pi$-coplanar phases in the latter region is fundamentally impossible from the symmetry-preserving finite-size calculations with fixed magnon number.} Therefore, we {also} perform a cluster mean-field plus scaling analysis for small spins $S\leq 3/2$. {The} obtained results{,} together with the previous large-$S$ analysis{, indicate} that the $\pi$-coplanar phase exists for any $S$ except for the classical limit ($S\rightarrow \infty$) and the existence range in $J/J_z$ is largest in the most quantum case of $S=1/2$. 
\end{abstract}
\maketitle
\section{\label{sec1}Introduction}
A strong magnetic field applied to a magnet forces the intrinsic spin moments to align along the field direction. Once {the magnetic saturation is reached}, the many-body state is given by a simple direct product of local maximum-spin ($S^z=S$) states (if the system is invariant under the spin rotation about the field axis, say the $z$ axis). Therefore, quantum-mechanical fluctuations are rather small in magnets near the saturation, compared to zero magnetic field. However, in strongly frustrated magnets, there can be 
a large number of possible magnetization processes that experience different magnetic phases but merge into the same saturated state as the magnetic field increases. This means that the energies of many different magnetic states are nearly degenerate in the vicinity of the saturation. Because of this, even small quantum fluctuations near the saturation could play a significant role in giving rise to exotic quantum phenomena, such as the spin nematic phase~\cite{shannon-06,zhitomirsky-10,svistov-10,sato-13}, {multi-$q$ phases}~\cite{kamiya-14,giacomo-14}, and nontrivial quantum criticality~\cite{jackeli-04,zhitomirsky-04}.

In this context of research, triangular-lattice antiferromagnets (TLAFs) have a long history as a promising model system for studying the interplay among frustration, quantum fluctuations, and magnetic fields~\cite{collins-97,starykh-15}. Recent advances in experiments have {allowed for preparing} a variety of quasi-two-dimensional (quasi-2D) TLAF materials and {accessing} their magnetic properties in strong magnetic fields up to the saturation field $H_s$~\cite{shirata-12,zhou-12,susuki-13,lee-14,yokota-14,koutroulakis-15,ma-16,sera-16,rawl-17}. {Since the seminal work by Kawamura and Miyashita~\cite{kawamura-85},} it has been known that the simplest model of TLAFs with isotropic Heisenberg interactions possesses an accidental continuous degeneracy of the classical ground state for finite magnetic fields. Past theoretical efforts have established that quantum (or thermal) fluctuations lift the classical degeneracy and select a magnetization process whose magnetization curve exhibits a plateau at one-third of the saturation magnetization~\cite{chubokov-91,yoshikawa-04,farnell-09,sakai-11,yamamoto-16,kawamura-85,seabra-11}. However, such a one-third magnetization plateau has been actually observed  only in a few TLAF materials~\cite{shirata-12,ono-03,fortune-09}. In real TLAF materials, one has to take into account some extra complexities such as Dzyaloshinskii-Moriya interaction, longer-range interactions, and some sort of anisotropy (single-ion, spin-exchange, spatial, etc.), which make further complications in determining the ground-state magnetic phase.

For instance, the layered TLAF materials such as Ba$_3$CoSb$_2$O$_9$~\cite{shirata-12,zhou-12,susuki-13,koutroulakis-15,ma-16,sera-16} and Ba$_3$CoNb$_2$O$_9$~\cite{lee-14,yokota-14} possess XXZ-type anisotropy in the spin-exchange interactions on each layer. The interactions between the spins on the single layer {are} modeled by the following XXZ Hamiltonian:
\begin{eqnarray}
\hat{\mathcal{H}}=
J\sum_{\langle i,j\rangle}\Big(\hat{S}_i^x\hat{S}_j^x+\hat{S}_i^y\hat{S}_j^y\Big)+J_z\!\sum_{\langle i,j\rangle}\hat{S}_i^z\hat{S}_j^z,
\label{hamiltonian}
\end{eqnarray}
where $\hat{\bm{S}}_i=(\hat{S}_i^x,\hat{S}_i^y,\hat{S}_i^z)$ is the spin operator with spin $S$ at site $i$ of the triangular lattice and the sum $\sum_{\langle i,j\rangle}$ runs over nearest-neighbor sites. The triangular-lattice XXZ model under a strong longitudinal magnetic field
\begin{eqnarray}
\hat{\mathcal{H}}_{\rm ext}=-H\sum_{i}\hat{S}^z_{i}\label{hz}
\end{eqnarray}
has recently received increasing attention since the latest theoretical studies~\cite{yamamoto-14,starykh-14} indicated the possibility for the emergence of a new magnetic phase, named the $\pi$-coplanar or $\Psi$ phase for $J>J_z>0$.

\begin{figure}[t]
\includegraphics[scale=0.35]{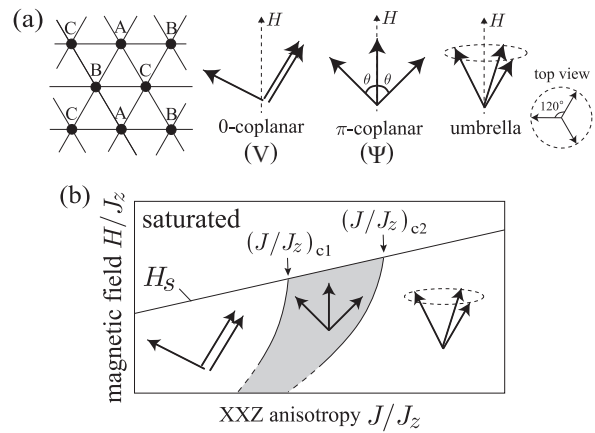}
\caption{\label{fig1}
Three magnetic phases predicted for the quantum triangular XXZ model in the presence of strong magnetic fields. (a) Spin moments on three sublattices A, B, and C in each phase are illustrated. (b) Schematic representation of the suggested phase diagram~\cite{yamamoto-14,starykh-14} in the vicinity of the saturation field $H_s=3(J+2J_z)S$.}
\end{figure}

In Ref.~\cite{yamamoto-14}, some of the {current} authors determined the ground-state phase diagram of the spin-1/2 XXZ model~(\ref{hamiltonian}) on the triangular lattice as a function of the XXZ anisotropy $J/J_z$ and the field strength $H$ with the use of the numerical cluster mean-field plus scaling (CMF+S) method. For strong magnetic fields, it was found that three different magnetic phases appear depending on the value of $J/J_z$: the so-called $0$-coplanar or {$V$} phase for small $J/J_z$, the umbrella phase for large $J/J_z$, and the $\pi$-coplanar or $\Psi$ phase in the intermediate region (see Fig.~\ref{fig1}). Whereas the former two already exist in the classical counterpart of the model~\cite{miyashita-86}, the latter emerges due to quantum-mechanical effects~\cite{yamamoto-14}. The critical values of the anisotropy $J/J_z$ at the phase transitions right below the saturation [$H=H_s-0^+$ with $H_s=3(J+2J_z)S$; see Fig.~\ref{fig1}(b)] were estimated as
\begin{eqnarray}
\begin{array}{lll}
(J/J_z)_{{\rm c}1}&=&1.588\\
(J/J_z)_{{\rm c}2}&=&2.220
\end{array}
\label{CMFSboundary}
\end{eqnarray}
within the CMF+S calculations for $S=1/2$~\cite{yamamoto-14}. For large spin values $S\gg 1$, Starykh $et$ $al$.~\cite{starykh-14} have estimated $(J/J_z)_{{\rm c}1}$ and $(J/J_z)_{{\rm c}2}$ based on the dilute Bose-gas expansion formalism, in which magnetic states in the vicinity of the saturation {are} described as Bose-Einstein condensations (BECs) of dilute magnons via the Holstein-Primakoff transformation~\cite{nikuni-95,batuev-84}. The values of $(J/J_z)_{{\rm c}1}$ and $(J/J_z)_{{\rm c}2}$ were determined at leading order in $1/S$ as
\begin{eqnarray}
\begin{array}{lll}
(J/J_z)_{{\rm c}1}&\approx& \left(1-0.45/S\right)^{-1}\\
&\approx & 1+0.45/S,\\
(J/J_z)_{{\rm c}2}&\approx& \left(1-0.53/S\right)^{-1}\\
&\approx & 1+0.53/S,
\end{array}
\label{largeS}\end{eqnarray}
for large S.

{More recently, in Ref.~\cite{giacomo-16}, some of the {current} authors calculated a quantitatively precise result for the coplanar-umbrella transition point at $H=H_s-0^+$ for arbitrary $S$, by treating the $1/S$ series exactly within the {dilute Bose-gas} framework~\cite{note}. This approach, however, makes it technically difficult to address the distinction between {the} 0-coplanar and $\pi$-coplanar states~\cite{nikuni-95,giacomo-16}.} {The transition point between the ({unspecified}) coplanar and umbrella phases, denoted by $(J/J_z)_{{\rm c}2^*}$, is obtained as $(J/J_z)_{{\rm c}2^*}=2.218$ for $S=1/2$, which is in excellent agreement with the $\pi$-coplanar/umbrella boundary given by the CMF+S calculation [$(J/J_z)_{{\rm c}2}=2.220$; Eq.~(\ref{CMFSboundary})]}.

The existence of the $\pi$-coplanar phase, {however}, is still under discussion in the case of small $S$, especially for $S=1/2$. Sellmann $et$ $al$. have performed an exact diagonalization (ED) analysis of Eq.~(\ref{hamiltonian}) with $S=1/2$ in the three-magnon 
sector (i.e., near the saturation)~\cite{sellmann-15} to reexamine the existence of the $\pi$-coplanar phase predicted in the CMF+S study~\cite{yamamoto-14}. Although they found three different parameter ranges of $J/J_z$ with different lowest eigenstates for $N=108$ spins or less, the intermediate region appeared to vanish when  the ED data were extrapolated to the limit of  infinite system size, $N\rightarrow \infty$. From this result, the authors of Ref.~\cite{sellmann-15} concluded that the ``$\pi$-coplanar'' phase is absent for $S=1/2$ in the thermodynamic limit. This clearly contradicts the CMF+S result~\cite{yamamoto-14} in which the $\pi$-coplanar region in the phase diagram gets {\it wider} as the cluster size increases. However, it should be noted that the conclusion by Sellmann $et$ $al$. relies on their speculation that the spurious phase that disappears in the thermodynamic limit would be the $\pi$-coplanar phase.

Finally, on the experimental side, Susuki $et$ $al$. have found that the spin-1/2 XXZ TLAF Ba$_3$CoSb$_2$O$_9$ has exhibited a magnetization anomaly at a strong transverse magnetic field $H\approx 0.7 H_s$~\cite{susuki-13}. This nontrivial anomaly has been thought to be due to the phase transition between the 0-coplanar and $\pi$-coplanar phases in the first report. However, its cause was later shown to be a first-order phase transition induced by small but nonvanishing interlayer coupling~\cite{koutroulakis-15,yamamoto-16,yamamoto-15}.

In this paper, we perform an ED and CMF+S study on the triangular-lattice XXZ model~(\ref{hamiltonian}) in longitudinal magnetic fields (\ref{hz}) to establish the magnetic phases that appear in the vicinity of the saturation field for small $S\leq 3/2$. The results also give {a resolution to} the contradiction between the CMF+S analysis~\cite{yamamoto-14} and the argument by Sellmann $et$ $al$.~{\cite{sellmann-15}}
regarding the existence of the $\pi$-coplanar phase. We carry out the ED calculations of {the model~(\ref{hamiltonian}) at $S=1/2$} on finite-size clusters of $N$ spins with periodic boundary condition. We {mainly} focus on {the} sector of the Hilbert space with only three down-spins ($\sum_i \hat{S}^z_i=N/2-3$) as in Ref.~\cite{sellmann-15} but consider {clusters of much larger size} up to $N=1296$. Several low-lying eigenvalues of the Hamiltonian are numerically computed {together with} the lowest eigenvalue as a function of the XXZ anisotropy parameter $J/J_z$. Moreover, we characterize each eigenstate according to the translational and point-group symmetries, and make further identification by calculating the overlap (inner product) with the finite-size coherent-state description of the candidate magnetic states (0-coplanar, $\pi$-coplanar, and umbrella states).

Our ED analysis {provides} the following results: First, we confirm {the presence of} three different $J/J_z$ ranges separated by two level crossings of the lowest and first-excited states {for relatively small $N$, which was found by Sellmann {\it et al.}, and that} the intermediate region actually vanishes when $N$ exceeds a certain size. Second, however, the lowest eigenstate that disappears in the thermodynamic limit is {\it not} the finite-size $\pi$-coplanar state, contrary to Sellmann $et$ $al$.'s speculation~\cite{sellmann-15}, but actually a chirally symmetric {superposition} of finite-size umbrella states. Third, the large-$J/J_z$ region is occupied by a chirally antisymmetric {superposition} of umbrella states. Last and most important, {in the small-$J/J_z$ region ({$J/J_z\lesssim 2.218$} at $N\rightarrow \infty$) the lowest level is doubly degenerate and} {its} {eigenstates correspond to a coplanar state.} {However, none of the available information allows us to distinguish the $\pi$-coplanar state from the 0-coplanar state.} Although we point out that the third and fourth low-lying eigenstates are crucial to lift the degeneracy, the distinction between {the 0-coplanar and $\pi$-coplanar states} is actually difficult since the third and fourth energy levels quickly {approach} {each other} as $N\rightarrow \infty$. The above-mentioned ED results deny the {claim in Ref.~\cite{sellmann-15}} {regarding} the nonexistence of the $\pi$-coplanar phase for $S=1/2$, and indicate the difficulty in distinguishing between {the 0-coplanar and $\pi$-coplanar states} in the symmetry-preserving finite-size calculations.

Furthermore, in order to complement the CMF+S study for $S=1/2$ in Ref.~\cite{yamamoto-14}, we also perform the CMF+S analysis for $S=1$ and $S=3/2$. In the CMF+S approach, the symmetry of the system is broken by self-consistent mean fields even on finite-size clusters, which enables us to distinguish between {the 0-coplanar and $\pi$-coplanar states}. The results for the transition points $(J/J_z)_{{\rm c}1}$ and $(J/J_z)_{{\rm c}2}$ just below the saturation field are extrapolated to the limit {of infinite} cluster size. Taking into consideration the results for $S=1$ and $S=3/2$ {and} the previous $S=1/2$ result [Eq.~(\ref{CMFSboundary})], we can see that the values of $(J/J_z)_{{\rm c}1}$ and $(J/J_z)_{{\rm c}2}$ are naturally approaching the $1/S$ estimation~\cite{starykh-14} given in Eq.~(\ref{largeS}) as $S$ increases. This strengthens our statement that the $\pi$-coplanar phase exists even for small $S$ down to $1/2$ and, moreover, the parameter ($J/J_z$) window to realize the $\pi$-coplanar phase is wider for smaller $S$.

The paper is organized as follows. In Sec.~\ref{sec2}, we present the ED calculations for $S=1/2$ on two series of clusters with different shapes. We show the level crossing between four low-lying eigenvalues and discuss the correspondence between the finite-size eigenstates and the expected magnetic phases in the thermodynamic limit. In Sec.~\ref{sec3}, some details about the CMF+S calculations for $S>1/2$ and the results for $S=1$ and $S=3/2$ are given. Section~\ref{sec4} is devoted to summary and conclusions.

\section{\label{sec2}Exact diagonalization analysis for $S=1/2$} 
\begin{figure}[b]
\includegraphics[scale=0.35]{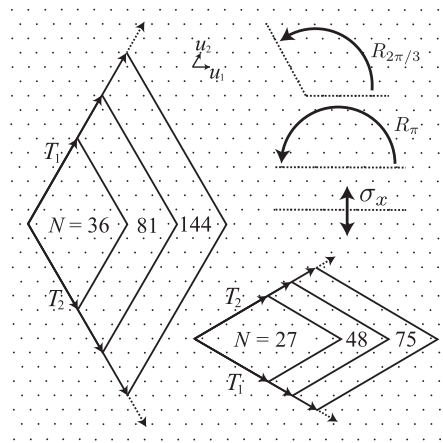}
\caption{\label{TriClusters}
{Two series of clusters that are considered in the present ED analysis.}}
\end{figure}
We first perform an ED analysis of the triangular-lattice spin-1/2 XXZ model~(\ref{hamiltonian}) on finite-size clusters of $N$ spins. We impose the periodic boundary condition defined by $\hat{\bm{S}}_{\bm{r}_i+\bm{T}_{1,2}}=\hat{\bm{S}}_{i}$ ($\bm{r}_i$ is the coordinates of site $i$) with the vectors 
\begin{eqnarray}
\bm{T}_1=l \bm{u}_1+m\bm{u}_2~~{\rm and}~~\bm{T}_2=-m \bm{u}_1+(l+m)\bm{u}_2,
\end{eqnarray}
where $\bm{u}_1=(1,0)$ and $\bm{u}_2=(1/2,\sqrt{3}/2)$~\cite{bernu-94}. To avoid the cluster shape dependence of the conclusion, we employ two series of clusters identified by $(l,m)=(3p,0)$ and $(l,m)=(2p,-p)$, respectively, with $p=1,2,\dots$, which are both compatible with the expected three-sublattice magnetic orders. The size of the clusters is given by $N=l^2+lm+m^2$ for both series, and specifically $N=9,36,81,144,\cdots$ and $N=3,12,27,48,\cdots$, respectively {(see Fig.~\ref{TriClusters})}. Unless specifically stated otherwise, the maximum size used in the calculations is {$N=1296$} for the former series and $N=1200$ for the latter. The data for $N\leq 12$ are not shown since the size is too small.

{Both} series possess the point-group symmetry under the transformations whose generators are the planar rotations $\mathcal{R}_{2\pi/3}$ and $\mathcal{R}_{\pi}$ of angles $2\pi/3$ and $\pi$, respectively, and the axial reflection $\sigma_x$ with respect to $\bm{u}_1$, as well as the translational symmetry $\mathcal{T}_{N}$. Thus the space group is {$\mathcal{G}_N=\mathcal{T}_{N}\rtimes \mathcal{C}_{6v}$}, which has $12N$ elements. Here, we mainly focus on the three-magnon sector of the Hilbert space where only three spins are down and the others are all up ($\sum_i \hat{S}^z_i=N/2-3$) as in Ref.~\cite{sellmann-15}. Thus the Hamiltonian matrix naively consists of {$_N {\rm C}_3\times {}_N {\rm C}_3$} components. For efficient calculations, we further divide the matrix into blocks of smaller dimensions according to the space-group symmetry, and numerically diagonalize each block matrix.

\subsection{\label{}Coherent-state description} 

Before showing the results of our ED calculations, {we} consider the coherent-state description of the candidate magnetic phases (0-coplanar, $\pi$-coplanar and umbrella states) . 
{Note that for strong fields one can exclude other phases from consideration, according to the dilute Bose-gas expansion~\cite{nikuni-95,giacomo-16}.} 
When the system size $N$ {increases} with {a} fixed number of magnons {$n$, the} density of magnons {$n/N$} goes to $0$. {Consequently}, the corresponding magnetic field $H$ approaches the saturation field $H_s$. In such a situation with dilute magnons, the magnetic order can be described by the Bose-Einstein condensation of magnons~\cite{nikuni-95,batuev-84}. {The} umbrella state is given by single BEC with either momentum $\bm{k}=\bm{Q}=(4\pi/3,0)$ or $-\bm{Q}$. The two options ($\pm\bm{Q}$) reflect the degeneracy with respect to the chirality. On the other hand, a double BEC with both momenta $\bm{k}=\bm{Q}$ and $\bm{k}=-\bm{Q}$ corresponds to coplanar states. The 0-coplanar and $\pi$-coplanar states are characterized by the relative phase $\phi$ between the condensates with $\bm{k}=\pm\bm{Q}$: $\phi=0$, $2\pi/3$, or $4\pi/3$ for 0-coplanar, and $\phi=\pi$, $5\pi/3$, or $\pi/3$ for $\pi$-coplanar. The three options for each reflect the remaining $Z_3$ symmetry with respect to the exchange of the three sublattices of the triangular lattice, A, B, and C in Fig.~\ref{fig1}. 

\begin{table}[tb]
(a)~0-coplanar order~$(m=2,3,\cdots)$
\begin{tabular}{cccc}
\hline
\hline
~~$n$~~~~& ~~$1$ ~~&~~$m$~~ \\ 
\hline 
~~irreps~~~~ & ~~$\Gamma_1,\Gamma_2$ ~~&~~$\Gamma_1,\Gamma_2,\Gamma_3$~~ \\
\hline 
\hline \\
\end{tabular}

(b)~$\pi$-coplanar order~$(m=1,2,\cdots)$
\begin{tabular}{cccc}
\hline
\hline
~~$n$~~~~& ~~$1$ ~~&~$2m$~~&~$2m+1$~~\\ 
\hline 
~~irreps~~~~ & ~~$\Gamma_1,\Gamma_2$~~ &~~$\Gamma_1,\Gamma_2,\Gamma_3$~~&~~$\Gamma_1,\Gamma_2,\Gamma_4$~~\\
\hline 
\hline \\
\end{tabular}

(c)~umbrella order~$(m=0,1,\cdots)$
\begin{tabular}{cccc}
\hline
\hline
~~$n$~~~~& ~$3m+1$ ~~&~$3m+2$~~&~$3m+3$~~ \\ 
\hline 
~~irreps~~~~ & ~~$\Gamma_1,\Gamma_2$~~ &~~$\Gamma_1,\Gamma_2$  ~~&~~$\Gamma_3,\Gamma_4$  ~~\\
\hline 
\hline 
\end{tabular}
\caption{\label{symmetry}Irreducible representations (irreps) of low-lying states characterizing the (a) 0-coplanar, (b) $\pi$-coplanar, and (c) umbrella phases, in each sector of $n$ magnons. The symbols $\Gamma_{1-4}$ are defined in Eq.~(\ref{irreps}).}
\end{table}

{In the regime of low magnon densities}, $n/N\rightarrow 0$, the magnetic orders may be described by the coherent states of dilute magnons: 
\begin{eqnarray}
\begin{array}{l}
|{\rm 0\mathchar`-coplanar}\rangle_{{\phi}_0} \\
~~~~\propto\displaystyle \exp\!\left[-\lambda \left(\hat{\mathcal{O}}_{ \bm{Q}}+e^{-i{\phi}_0}\hat{\mathcal{O}}_{- \bm{Q}}\right)\right]|{\rm sat}\rangle\\
\\
|\pi{\rm \mathchar`-coplanar}\rangle_{{\phi}_0} \\
~~~~\propto\displaystyle \exp\!\left[-\lambda \left(\hat{\mathcal{O}}_{ \bm{Q}}-e^{-i{\phi}_0}\hat{\mathcal{O}}_{- \bm{Q}}\right)\right]|{\rm sat}\rangle\\
\\
|{\rm umbrella}\rangle_\pm \propto  \displaystyle \exp\!\left(-\lambda \hat{\mathcal{O}}_{\pm\bm{Q}}\right)|{\rm sat}\rangle, 
\end{array}
\end{eqnarray}
where ${\phi}_0=0$, $2\pi/3$, or $4\pi/3$; $\lambda$ is, in general, a complex number; and $|{\rm sat}\rangle$ is the magnetically saturated state given by a direct product of local spin-up states. The magnon ``creation'' operators $\hat{\mathcal{O}}_{ \bm{Q}}$ and $\hat{\mathcal{O}}_{- \bm{Q}}$ are given by
\begin{eqnarray}
\hat{\mathcal{O}}_{\pm \bm{Q}}= \frac{1}{N}\sum_i \hat{S}^-_i e^{ \pm i\bm{Q}\cdot \bm{r}_i}
\end{eqnarray}
in the spin language. 
Therefore, in the sector of $n$ magnons, the {states corresponding to} each magnetic order should be given by
\begin{eqnarray}
\begin{array}{lll}
|{\rm 0\mathchar`-coplanar}\rangle^{(n)}_{{\phi}_0}&\!\propto\!& \displaystyle \left(e^{i{\phi}_0/2}\hat{\mathcal{O}}_{ \bm{Q}}+e^{-i{\phi}_0/2}\hat{\mathcal{O}}_{- \bm{Q}}\right)^n|{\rm sat}\rangle\\
&&\\
|\pi{\rm \mathchar`-coplanar}\rangle^{(n)}_{{\phi}_0}&\!\propto\!& \displaystyle \left(e^{i{\phi}_0/2}\hat{\mathcal{O}}_{ \bm{Q}}-e^{-i{\phi}_0/2}\hat{\mathcal{O}}_{- \bm{Q}}\right)^n|{\rm sat}\rangle\\
&&\\
|{\rm umbrella}\rangle^{(n)}_\pm&\!\propto\!&  \displaystyle \hat{\mathcal{O}}_{\pm\bm{Q}}^n|{\rm sat}\rangle 
\end{array}\label{statesN}
\end{eqnarray}
for $N\rightarrow \infty$.

The operators $\hat{\mathcal{O}}_{ \bm{Q}}$ and $\hat{\mathcal{O}}_{- \bm{Q}}$ are transformed by the point-group symmetry transformations as 
\begin{eqnarray}
\begin{array}{rc}
\mathcal{R}_{2\pi/3}:&~~\hat{\mathcal{O}}_{ \bm{Q}}\rightarrow \hat{\mathcal{O}}_{ \bm{Q}}~{\rm and}~\hat{\mathcal{O}}_{- \bm{Q}}\rightarrow \hat{\mathcal{O}}_{ -\bm{Q}},\\
\mathcal{R}_{\pi}:&~~\hat{\mathcal{O}}_{ \bm{Q}}\rightarrow \hat{\mathcal{O}}_{ -\bm{Q}}~{\rm and}~\hat{\mathcal{O}}_{- \bm{Q}}\rightarrow \hat{\mathcal{O}}_{ \bm{Q}},\\
\sigma_x:&~~\hat{\mathcal{O}}_{ \bm{Q}}\rightarrow \hat{\mathcal{O}}_{ \bm{Q}}~{\rm and}~\hat{\mathcal{O}}_{- \bm{Q}}\rightarrow \hat{\mathcal{O}}_{ -\bm{Q}}.
\end{array}\label{tra}
\end{eqnarray}
From Eqs.~(\ref{statesN}) and (\ref{tra}), the irreducible representations (irreps) characterizing each magnetic order are listed in Table~\ref{symmetry}. 
Here, $\Gamma_{1}$-$\Gamma_{4}$ are defined by
\begin{eqnarray}
\begin{array}{rl}
\Gamma_{1}=&[{\bm k}={\bm Q},~\mathcal{R}_{2\pi/3}=1,~\sigma_x=1],\\
\Gamma_{2}=&[{\bm k}={-\bm Q},~\mathcal{R}_{2\pi/3}=1,~\sigma_x=1],\\
\Gamma_{3}=&[{\bm k}={\bm 0},~\mathcal{R}_{2\pi/3}=1,~\mathcal{R}_{\pi}=1,~\sigma_x=1],\\
\Gamma_{4}=&[{\bm k}={\bm 0},~\mathcal{R}_{2\pi/3}=1,~\mathcal{R}_{\pi}=-1,~\sigma_x=1].
\end{array}\label{irreps}
\end{eqnarray}
As seen from Table~\ref{symmetry}, {the three magnetic orders of interest are distinguishable by the irreps only in the sectors of $n=2m+1$ magnons ($m=1,2,\cdots$), of which we mainly focus on the minimum one $n=3$ ($\sum_i \hat{S}^z_i=N/2-3$). Since the calculation cost grows quickly with the magnon number $n$, we will push the analysis to $n=5,7$ only for several values of $J/J_z$. }

\subsection{\label{sec2b}Low-lying eigenstates in the three-magnon sector} 
\begin{figure}[t]
\includegraphics[scale=0.7]{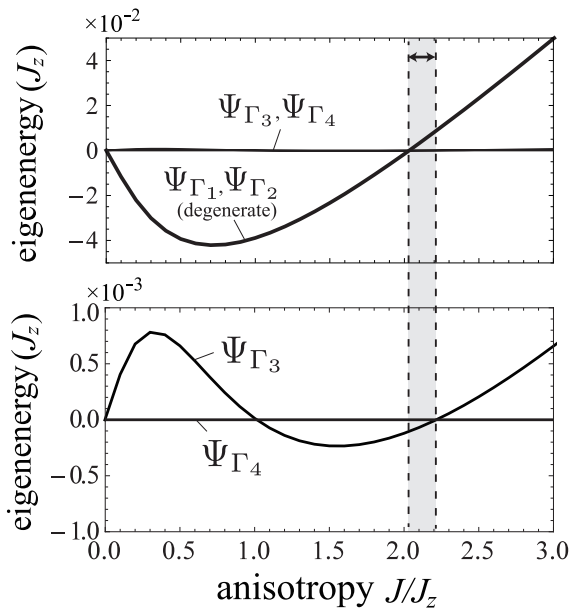}
\caption{\label{EnergyDiagram36}
The low-lying eigenenergies of the three-magnon sector for $N=36$ as a function of $J/J_z$. We plot the eigenenergies of the $\Psi_{\Gamma_{1}}$, $\Psi_{\Gamma_{2}}$, and $\Psi_{\Gamma_{3}}$ states measured from that of $\Psi_{\Gamma_{4}}$. The lower panel is the enlarged view showing the energy difference between $\Psi_{\Gamma_{3}}$ and $\Psi_{\Gamma_{4}}$. In the shaded region marked by the double-headed arrow, the $\Psi_{\Gamma_{3}}$ state has the lowest energy. }
\end{figure}
In Fig.~\ref{EnergyDiagram36}, we show the four low-lying eigenenergies for $N=36$ in the three-magnon sector ($n=3$ or $\sum_i\hat{S}^z_i=N/2-3$) as a function of $J/J_z$. The higher eigenstates are clearly separated from them. Thus, the four low-lying eigenstates are possible candidates for {quasidegenerate joint states}~\cite{bernu-94}. Each eigenstate is labeled by the irreps given in Eq.~(\ref{irreps}), e.g., $\Psi_{\Gamma_1}$. Note that $\Psi_{\Gamma_1}$ and $\Psi_{\Gamma_2}$ are completely degenerate {since the system has a trivial inversion symmetry of the wavevector, $\bm{k}\leftrightarrow -\bm{k}$}. On the other hand, $\Psi_{\Gamma_3}$ and $\Psi_{\Gamma_4}$ are not degenerate, but the energy difference is much smaller than that between $\Psi_{\Gamma_{1,2}}$ and $\Psi_{\Gamma_{3,4}}$.

\begin{figure}[tb]
\includegraphics[scale=0.7]{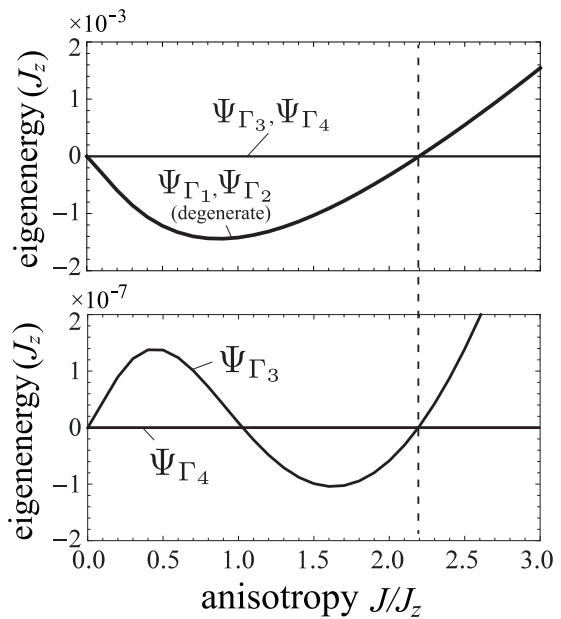}
\caption{\label{EnergyDiagram324}
Same as in Fig.~\ref{EnergyDiagram36}, but for $N=324$. There is no $J/J_z$ range where the {$\Gamma_{3}$} state has the lowest energy. }
\end{figure}
\begin{figure}[t]
\includegraphics[scale=0.55]{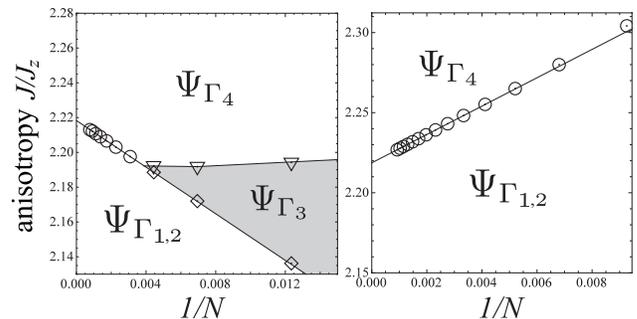}
\caption{\label{Crossing}
{Lowest eigenstate diagram of $1/N$ versus $J/J_z$. The eigenenergy-crossing points between the $\Psi_{\Gamma_{1,2}}$ and $\Psi_{\Gamma_{4}}$ states (marked by the circles), between the $\Psi_{\Gamma_{1,2}}$ and $\Psi_{\Gamma_{3}}$ states (diamonds) and between the $\Psi_{\Gamma_{3}}$ and $\Psi_{\Gamma_{4}}$ states (inverted triangles) are shown. The boundary lines are just a guide for the eye. The left (right) panel shows the data for the series of {$N=36,81,\cdots,1296$} ($N=27,48,\cdots,1200$).} }
\end{figure}
When the anisotropy parameter $J/J_z$ is increased from easy-axis to easy-plane, the lowest eigenstate is changed from the degenerate $\Psi_{\Gamma_{1,2}}$ to $\Psi_{\Gamma_3}$, then to $\Psi_{\Gamma_4}$ through two eigenenergy crossings. However, {as the system size $N$ increases, the range where $\Psi_{\Gamma_3}$ has the lowest energy shrinks and it disappears at $N\geq 324$}, in the case of the cluster series of {$N=36,81,\cdots,1296$}. In Fig.~\ref{EnergyDiagram324}, we show the eigenenergy diagram for $N=324$. Moreover, in the case of the cluster series of $N=27,48,\cdots,1200$, there is no window where $\Psi_{\Gamma_3}$ has the lowest energy even at $N=27$. The eigenenergy-crossing points are summarized in Fig.~\ref{Crossing}. 
In either case of the two cluster series, the degenerate $\Psi_{\Gamma_{1,2}}$ state is the lowest eigenstate for {$J/J_z\lesssim 2.218$}  while the $\Psi_{\Gamma_{4}}$ is the lowest for {$J/J_z\gtrsim 2.218$} in the thermodynamic limit.

\subsection{\label{sec2c}Identification of the low-lying eigenstates} 
In Ref.~\cite{sellmann-15}, Sellmann $et$ $al$. supposed that the $\Psi_{\Gamma_{1,2}}$, $\Psi_{\Gamma_3}$, and $\Psi_{\Gamma_4}$ states corresponded to the 0-coplanar, $\pi$-coplanar, and umbrella orders, respectively, and thus concluded that the $\pi$-coplanar phase was not stabilized as the ground state in the thermodynamic limit for $S=1/2$. However, it should be noted that their argument lacked {a precise} identification of the eigenstates. 
In the following, we identify each low-lying eigenstate {on the basis of} the space-group symmetry (\ref{irreps}) and, more crucially, by calculating the overlap with the coherent states (\ref{statesN}) of the expected magnetic orders. {Note that a similar identification approach based on the coherent-state description of long-range orders has been used for the study of a multiple-spin exchange model and has successfully identified the emergence of a spin nematic order~\cite{momoi-12}}.

From Eq.~(\ref{statesN}), the finite-size 0-coplanar, $\pi$-coplanar, and umbrella states are given in the three-magnon ($n=3$) sector by
\begin{eqnarray}
\begin{array}{lll}
|{\rm 0\mathchar`-coplanar}\rangle^{(3)}_{\phi_0}&\propto&\displaystyle 3e^{i\phi_0/2}|\Psi^*_{\Gamma_1}\rangle+3e^{-i\phi_0/2}|\Psi^*_{\Gamma_2}\rangle\\
&& +{\cos \left(3\phi_0/2\right)} |\Psi^*_{\Gamma_3}\rangle\\
&&\\
|\pi{\rm \mathchar`-coplanar}\rangle^{(3)}_{\phi_0}&\propto&\displaystyle 3e^{i\phi_0/2}|\Psi^*_{\Gamma_1}\rangle-3e^{-i\phi_0/2}|\Psi^*_{\Gamma_2}\rangle\\
&&-{\cos \left(3\phi_0/2\right)} |\Psi^*_{\Gamma_4}\rangle\\
&&\\
|{\rm umbrella}\rangle^{(3)}_{\pm}&\propto&|\Psi^*_{\Gamma_3}\rangle\pm |\Psi^*_{\Gamma_4}\rangle,
\end{array}\label{statesN*}
\end{eqnarray}
where
\begin{eqnarray}
\begin{array}{lll}
|\Psi^*_{\Gamma_1}\rangle&\equiv&\displaystyle \hat{\mathcal{O}}_{ \bm{Q}}^2\hat{\mathcal{O}}_{- \bm{Q}}|{\rm sat}\rangle,\\
&&\\
|\Psi^*_{\Gamma_2}\rangle&\equiv&\displaystyle  \hat{\mathcal{O}}_{ \bm{Q}}\hat{\mathcal{O}}_{- \bm{Q}}^2|{\rm sat}\rangle,\\
&&\\
|\Psi^*_{\Gamma_3}\rangle&\equiv&\displaystyle  \left(\hat{\mathcal{O}}_{ \bm{Q}}^3+\hat{\mathcal{O}}_{- \bm{Q}}^3\right)|{\rm sat}\rangle,\\
&&\\
|\Psi^*_{\Gamma_4}\rangle&\equiv&\displaystyle  \left(\hat{\mathcal{O}}_{ \bm{Q}}^3-\hat{\mathcal{O}}_{- \bm{Q}}^3\right)|{\rm sat}\rangle.
\end{array}\label{st}
\end{eqnarray}
The subscript $\Gamma_{m}$ {indicates} the irrep characterizing the state $\Psi^*_{\Gamma_m}$, which can be easily obtained from Eq.~(\ref{tra}).

Since the {symmetry} is not spontaneously broken in finite-size systems, the eigenstates of the Hamiltonian~(\ref{hamiltonian}) must belong to one of the irreps of the space group {$\mathcal{G}_N=\mathcal{T}_{N}\rtimes \mathcal{C}_{6v}$}. Therefore, for example, the finite-size umbrella state appears only in the form of the chirally symmetric combination
\begin{eqnarray*}
|{\rm umbrella}\rangle^{(3)}_{+}+|{\rm umbrella}\rangle^{(3)}_{-} \propto |\Psi^*_{\Gamma_3}\rangle
\end{eqnarray*}
or the chirally antisymmetric combination 
\begin{eqnarray*}
|{\rm umbrella}\rangle^{(3)}_{+}-|{\rm umbrella}\rangle^{(3)}_{-}\propto |\Psi^*_{\Gamma_4}\rangle. 
\end{eqnarray*}
In a similar fashion, the $Z_3$ symmetry with respect to $\phi_0$ in the finite-size 0-coplanar (respectively $\pi$-coplanar) state is not broken for finite $N$, and the basis symmetry-preserving states $\Psi^*_{\Gamma_1}$, $\Psi^*_{\Gamma_2}$, and $\Psi^*_{\Gamma_3}$ (respectively $\Psi^*_{\Gamma_1}$, $\Psi^*_{\Gamma_2}$, and $\Psi^*_{\Gamma_4}$) appear as separate eigenstates.

\begin{figure}[tb]
\includegraphics[scale=0.6]{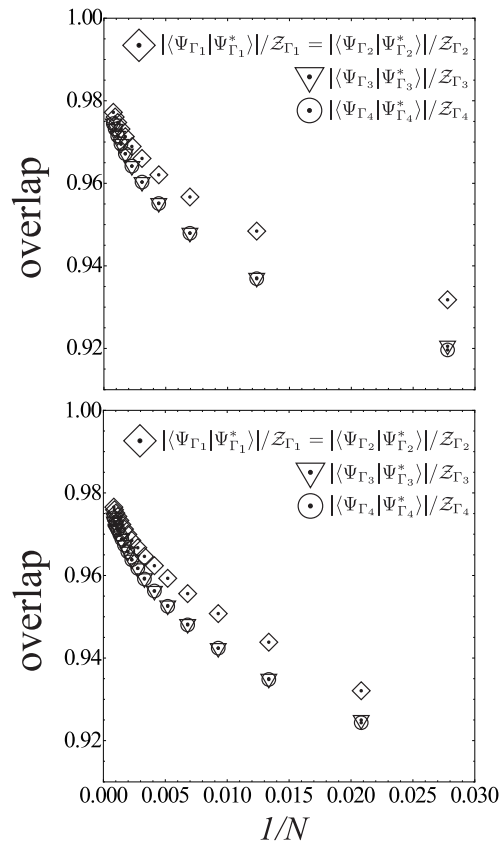}
\caption{\label{overlap}
Overlaps between $\Psi_{\Gamma_{m}}$ and $\Psi^*_{\Gamma_{m}}$ as a function of $1/N$ at $J/J_z=1$. The top and bottom panels show the data for the series of $N=36,81,\cdots,1296$ and $N=27,48,\cdots,1200$, respectively. All data {seem} to approach $1$ as $N\rightarrow\infty$. }
\end{figure}
From the correspondence of the symmetry property, it can be naturally expected that the eigenstates calculated by the exact diagonalization, $\Psi_{\Gamma_{1-4}}$ in Sec.~\ref{sec2b}, correspond to $\Psi^*_{\Gamma_{1-4}}$ of Eq.~(\ref{st}) in the thermodynamic limit, i.e., $|\Psi_{\Gamma_{m}}\rangle\rightarrow |\Psi^*_{\Gamma_{m}}\rangle$ ($N\rightarrow \infty$; $m=1,2,3,4$). We confirm this expectation by numerically calculating the overlap (inner-product) between $|\Psi_{\Gamma_{m}}\rangle$ and the corresponding $|\Psi^*_{\Gamma_{m}}\rangle$ as a function of $1/N$. Figure~\ref{overlap} shows an example of the results. {The overlaps $\langle \Psi_{\Gamma_{m}}|\Psi^*_{\Gamma_{m}}\rangle/\mathcal{Z}_{\Gamma_{m}}$, where $\mathcal{Z}_{\Gamma_{m}}\equiv \sqrt{\displaystyle \langle \Psi_{\Gamma_{m}}|\Psi_{\Gamma_{m}}\rangle\langle \Psi^*_{\Gamma_{m}}|\Psi^*_{\Gamma_{m}}\rangle}$ is the normalization factor, approach 1 as $N\rightarrow \infty$ as expected for all $m=1,2,3,4$}.

Having the above facts in mind, let us {look again at} the results of Figs.~\ref{EnergyDiagram36},~\ref{EnergyDiagram324}, and~\ref{Crossing}: (i) There is no $J/J_z$ range where the $\Psi_{\Gamma_{3}}$ state has the lowest energy in the thermodynamic limit as shown in Fig.~\ref{Crossing}, which, however, does {\it not} necessarily mean the absence of the $\pi$-coplanar phase for $S=1/2$. This is because $\Psi_{\Gamma_{3}}$ is not the $\pi$-coplanar state but the chirally symmetric combination of finite-size umbrella states (unlike the interpretation of Ref.~\cite{sellmann-15}). (ii) The double degeneracy of the lowest-energy eigenstates $\Psi_{\Gamma_{1}}$ and $\Psi_{\Gamma_{2}}$ in the region of {$J/J_z\lesssim 2.218$} ($N\rightarrow \infty$) {should} correspond to a coplanar phase according to Eq.~(\ref{statesN*}), although one cannot {identify only from the lowest eigenstate which is the ground state at the thermodynamic limit, the 0-coplanar or $\pi$-coplanar state}. (iii) For {$J/J_z\gtrsim 2.218$}, the lowest eigenstate is $\Psi_{\Gamma_{4}}$, which is the chirally antisymmetric combination of finite-size umbrella states. Therefore, the threshold value {$J/J_z\approx 2.218$} may indicate the coplanar-umbrella transition point just below the saturation field. The value is indeed in good agreement with the semianalytical value $(J/J_z)_{{\rm {\rm c}2}^*}= 2.218$ {of the dilute Bose-gas expansion} for the coplanar-umbrella transition point~\cite{giacomo-16}.

\subsection{\label{}0-coplanar or $\pi$-coplanar} 
As seen from the discussion of Sec.~\ref{sec2c}, only the consideration of the lowest eigenstate is not {sufficient to determine which state is selected as the ground state in the region of $J/J_z\lesssim 2.218$, 0-coplanar or $\pi$-coplanar, because} the degenerate eigenstates $\Psi_{\Gamma_{1,2}}$ are common for general coplanar orders [see Eq.~(\ref{statesN*})]. {Hence}, one needs to take higher eigenstates into consideration.

\begin{figure}[tb]
\includegraphics[scale=0.45]{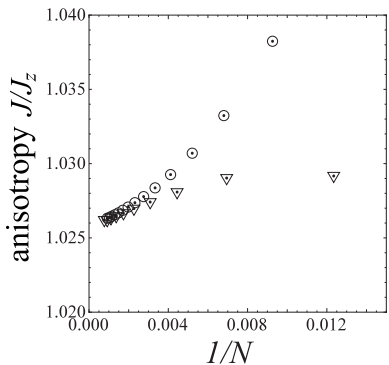}
\caption{\label{Crossing2}
The energy-crossing point around $J/J_z\sim 1$ between the eigenstates $\Psi_{\Gamma_{3}}$ and $\Psi_{\Gamma_{4}}$ as a function of $1/N$. The circles and inverted triangles show the data for the series of {$N=36,81,\cdots,1296$} and $N=27,48,\cdots,1200$, respectively. }
\end{figure}

As seen in the ED data of Fig.~\ref{EnergyDiagram324}, the energy of $\Psi_{\Gamma_{4}}$ is smaller than that of $\Psi_{\Gamma_{3}}$ for $J/J_z\lesssim 1.03$ while it is opposite for $1.03\lesssim J/J_z\lesssim 2.2$. This feature does not depend on the system size or the cluster shape (see Fig.~\ref{Crossing2}). According to Eq.~(\ref{statesN*}), if the eigenstates $\Psi_{\Gamma_{1,2}}$ and $\Psi_{\Gamma_{3}}$ (respectively $\Psi_{\Gamma_{1,2}}$ and $\Psi_{\Gamma_{4}}$) collapse to the ground state in the the thermodynamic limit, the 0-coplanar (respectively $\pi$-coplanar) ground state would be formed as a result of the spontaneous $U(1)\times Z_3$ symmetry breaking. Therefore, naively considering the ED data, one might conclude that the $\pi$-coplanar phase emerges for $0< J/J_z\lesssim 1.03$ while the $0$-coplanar phase is stabilized for $1.03\lesssim J/J_z\lesssim 2.22$.

However, this conclusion {is in clear contrast} with the widely accepted consensus that the ground state of the quantum isotropic Heisenberg model ($J/J_z=1$) on the triangular lattice exhibits the 0-coplanar ($V$) magnetic order for strong magnetic fields~\cite{chubokov-91}. This discrepancy can be understood in the following way. In the ED calculations, we consider {a} sector of the Hilbert space with a fixed number of magnons ($n=3$ in this paper). Therefore, the ``thermodynamic limit'' $N\rightarrow \infty$ means the ``dilute-magnon limit'' $n/N\rightarrow 0$ (i.e., $H\rightarrow H_s-0^+$) at the same time. It should be recalled that in such a limit the 0-coplanar, $\Psi$-coplanar, and umbrella states become all degenerate and merge into the {saturated} state at $H=H_s$. Thus, the finite-size ED analysis can distinguish {between} the 0-coplanar and $\pi$-coplanar states only when the merging of the two states as $H\rightarrow H_s-0^+$ is slower than the collapse of the basis eigenstates in each state ($\{\Psi_{\Gamma_{1,2}}$, $\Psi_{\Gamma_{3}}\}$ or $\{\Psi_{\Gamma_{1,2}}$, $\Psi_{\Gamma_{4}}\}$) to the ground state as $N\rightarrow \infty$.

\begin{figure}[tb]
\includegraphics[scale=0.45]{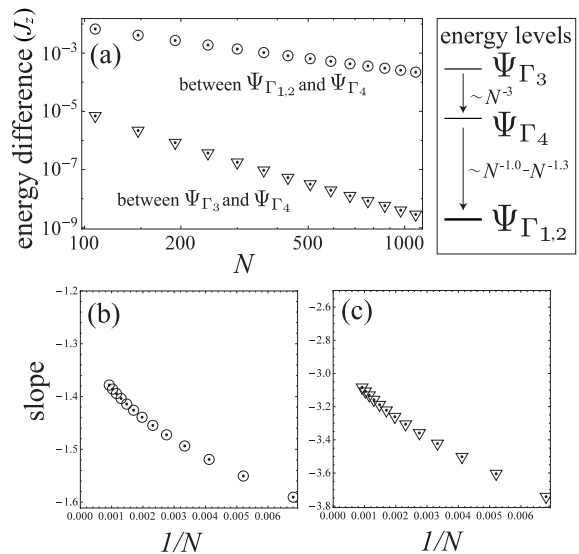}
\caption{\label{Scaling05}
(a) Log-log plots of the eigenenergy differences between $\Psi_{\Gamma_{1,2}}$ and $\Psi_{\Gamma_{4}}$ (circles) and between $\Psi_{\Gamma_{3}}$ and $\Psi_{\Gamma_{4}}$ (inverted triangles) as a function of the system size $N$. The {slope} of each pair of the two neighboring data points are shown in (b) for the former and (c) for the latter as a function of $1/N$. The right-hand panel of (a) is the schematic diagram of the energy levels in the range of $0<J/J_z\lesssim 1.03$. We show the data at $J/J_z=0.5$ for the cluster series of $N=27,48,\cdots, 1200$ as an example.
}
\end{figure}
In Fig.~\ref{Scaling05}, we show a typical example of the finite-size scaling of the eigenenergy differences between $\Psi_{\Gamma_{1,2}}$ and $\Psi_{\Gamma_{4}}$ and between $\Psi_{\Gamma_{3}}$ and $\Psi_{\Gamma_{4}}$. The energy of the eigenstate $\Psi_{\Gamma_{4}}$ approaches the lowest $\Psi_{\Gamma_{1,2}}$ level, and the difference goes to zero as $\sim N^{-\alpha}$ with $\alpha\approx 1.0$-$1.3$, which seems slightly faster than the Nambu-Goldstone mode expected to collapse as $\sim N^{-1}$. However, at the same time, the $\Psi_{\Gamma_{3}}$ level also collapses into the ground state. More importantly, the merging of $\Psi_{\Gamma_{3}}$ into $\Psi_{\Gamma_{4}}$ is much faster ($\sim N^{-3}$) than the collapse of $\Psi_{\Gamma_{4}}$ and $\Psi_{\Gamma_{1,2}}$. This means that it is fundamentally impossible to distinguish {between} the 0-coplanar and $\pi$-coplanar states from the finite-size eigenstates with {a} fixed number of magnons.

\subsection{\label{}Analysis with a fixed density of magnons} 
As seen above, the finite-size ED analysis with a fixed number of magnons is inadequate for discussing the relative angles among the three sublattice moments in coplanar states. The point is that both 0-coplanar and $\pi$-coplanar states merge into the saturated state as $n/N\rightarrow 0$ ($H\rightarrow H_s-0^+$) and become indistinguishable from each other. 
One may avoid this issue by fixing the {\it density} of magnons, $n/N$, instead of the {\it number} of magnons $n$ when taking the thermodynamic limit $N\rightarrow \infty$. In the magnetic phases supposed here (0-coplanar, $\pi$-coplanar, or umbrella), the finite-size gap of the Nambu-Goldstone mode (linear in the wavelength) is expected to scale as $\sim N^{-1/2}$ at fixed magnon density~\cite{bernu-94}. Therefore, the magnetic phase that appears in the thermodynamic limit is determined by the finite-size eigenstates that collapse into the ground state faster than $N^{-1/2}$. Each candidate magnetic order is identified by the eigenstates with the irreps listed in Table~\ref{symmetry}.

In order to take the {thermodynamic} limit at a fixed magnon density, it is required to perform the ED calculations with different values of magnon number $n$; however, the computable system size decreases rapidly as $n$ increases. Moreover, according to the irreps listed in Table~\ref{symmetry}, only the Hilbert space sector with an odd number of magnons $n=2m+1$ ($m=1,2,\cdots$) can distinguish between the 0-coplanar and $\pi$-coplanar states. Therefore, we consider the magnon numbers $n=3$, 5, and 7 and the cluster series of $N=27,48,\cdots$. The maximum system size computed in the present work for $n=7$ is $N=108$, and thus we shall 
{take a small but fixed magnon density, namely $n/N=7/108\approx 0.065$.} According to Table~\ref{symmetry}, the 0-coplanar (respectively $\pi$-coplanar) state is expected to be formed in the thermodynamic limit when the $\Psi_{\Gamma_3}$ (respectively $\Psi_{\Gamma_4}$) state collapses into the lowest $\Psi_{\Gamma_{1,2}}$ state faster than $N^{-1/2}$ and is separated from the $\Psi_{\Gamma_4}$ (respectively $\Psi_{\Gamma_3}$) state in a distinguishable fashion.

\begin{figure}[tb]
\includegraphics[scale=0.5]{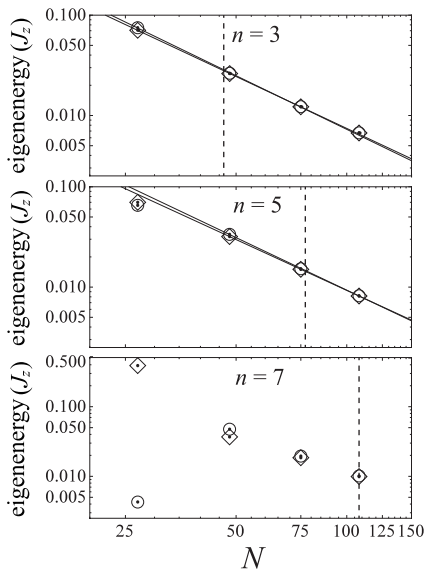}
\caption{\label{357mag}
Log-log plots of the eigenenergies of $\Psi_{\Gamma_{3}}$ (circles) and $\Psi_{\Gamma_{4}}$ (diamonds) measured from that of the lowest eigenstate $\Psi_{\Gamma_{1,2}}$ as a function of the system size $N$ at $J/J_z=0.5$. We show the data for different magnon numbers ($n=3$, 5, and 7). The lines for $n=3$ ($n=5$) are linear fitting functions of the $N=27,48,75$ ($N=48,75,108$) data for each data series. The vertical dashed lines mark the point at which the magnon density $n/N=7/108\approx 0.065$. 
}
\end{figure}
In Fig.~\ref{357mag}, we show the size dependence of the eigenenergies of the $\Psi_{\Gamma_{3}}$ and $\Psi_{\Gamma_{4}}$ states measured from that of the lowest eigenstate $\Psi_{\Gamma_{1,2}}$ at $J/J_z=0.5$ (as an example) for $n=3$, 5, and 7, respectively. The system size $N$ can take only specific values. Therefore, we employ the following interpolation scheme in order to fix the magnon density to $n/N=7/108\approx 0.065$ for $n=3$ and 5: Using the log-log plots shown in Fig.~\ref{357mag}, we perform a linear least-squares fitting for the data of the three system sizes neighboring to $N= n/0.065$, that is, $N=28,48,75$ for $n=3$ and $N=48,75,108$ for $n=5$. From the fitting functions, we obtain the interpolation values of the eigenenergy differences that correspond to the magnon density $n/N=0.065$. The same procedure is also performed for other values of $J/J_z$.

\begin{figure}[tb]
\includegraphics[scale=0.5]{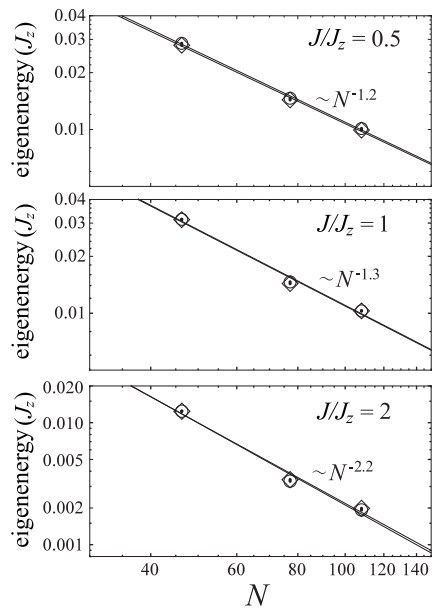}
\caption{\label{densityfixed}
Log-log plots of the eigenenergies of $\Psi_{\Gamma_{3}}$ (circles) and $\Psi_{\Gamma_{4}}$ (diamonds) measured from that of the lowest eigenstate $\Psi_{\Gamma_{1,2}}$ as a function of the system size $N$ at the fixed magnon density $n/N=0.065$. We show the data for different anisotropy parameters ($J/J_z=0.5$, 1, and 2). The lines are a linear fitting function for each data series.
}
\end{figure}
In Fig.~\ref{densityfixed}, we show the size dependence of the eigenenergies of the $\Psi_{\Gamma_{3}}$ and $\Psi_{\Gamma_{4}}$ states measured from that of the lowest eigenstate $\Psi_{\Gamma_{1,2}}$ at $J/J_z=0.5$, 1, and 2, after refining the data such that the magnon density is fixed to $n/N=0.065$. 
Contrary to expectation, we find no clear splitting of the $\Psi_{\Gamma_{3}}$ and $\Psi_{\Gamma_{4}}$ levels for any case, although it should, in principle, be possible to distinguish between 0-coplanar and $\pi$-coplanar states from the analysis with a fixed magnon density as mentioned above. This might be attributed to the fact that the number of magnons $n\leq 7$ and, correspondingly, the size of the system $N\leq 108$ are still too small to see a clear separation of the $\Psi_{\Gamma_{3}}$ and $\Psi_{\Gamma_{4}}$ levels{; also, we might have to reconsider more carefully the mechanism which leads to the generation of the symmetry broken state in the thermodynamic limit.}

Surprisingly, even for the isotropic Heisenberg model ($J/J_z=1$), although it is {widely believed} that the 0-coplanar phase is formed for strong magnetic fields, one cannot rigorously specify the coplanar phase, 0 or $\pi$, from the ``exact'' calculations for finite-size systems at the stage of the current numerical capabilities. Further developments of computer performance and numerical techniques are anticipated to solve this open fundamental problem regarding the spontaneous symmetry breaking in frustrated systems.

\section{\label{sec3}Cluster mean-field theory with cluster-size scaling for $S\leq 3/2$} 

{We found out in Sec.~\ref{sec2} that it is difficult to address the issue of the distinction between the 0-coplanar and $\pi$-coplanar states by using the symmetry-preserving ED analysis on finite systems}. 
{In the case of analytical studies with the {dilute Bose-gas} expansion~\cite{nikuni-95,giacomo-16},} 
the calculation of sixth-order corrections {to the ground-state energy} is required to address the distinction between {the 0-coplanar and $\pi$-coplanar states}, but it is {also} technically difficult.

One promising approach to address the problem is the CMF+S method~\cite{yamamoto-16,yamamoto-14,yamamoto-15,yamamoto-12-2}. This method has been applied to the triangular-lattice XXZ model~(\ref{hamiltonian}) for $S=1/2$ and has successfully {produced} the ground-state phase diagram including the transition between the 0-coplanar and $\pi$-coplanar phases~\cite{yamamoto-14}. In the CMF+S method, although one deals with interacting spins on a finite-size cluster as in the ED analysis, the correlation effects from the spins outside the cluster are also treated as effective magnetic fields (mean fields) acting on the edge sites of the cluster.

The many-body problem with the Hamiltonian~(\ref{hamiltonian}) in the presence of external magnetic fields (\ref{hz}) is replaced by an $N_C$-body problem described by the cluster Hamiltonian
\begin{eqnarray}
\hat{\mathcal{H}}_C&=&
J\sum_{\langle i,j\rangle\in C}^{N_B}\Big(\hat{S}_i^x\hat{S}_j^x+\hat{S}_i^y\hat{S}_j^y\Big)+J_z\!\sum_{\langle i,j\rangle\in C}^{N_B}\hat{S}_i^z\hat{S}_j^z\nonumber\\
&&-H\sum_{i\in C}^{N_C}\hat{S}_i^z-\sum_{i\in {\partial C}}{\bm h}_i^{\rm MF}\cdot\hat{\bm S}_i,
\label{chamiltonian}
\end{eqnarray}
where $N_B$ is the number of nearest-neighbor bonds inside the cluster $C$ and $\partial C$ are the edge sites of the cluster. The mean-field decoupling $\hat{S}_i^\alpha \hat{S}_j^\alpha\rightarrow \langle \hat{S}_i^\alpha\rangle \hat{S}_j^\alpha+\langle\hat{S}_j^\alpha\rangle \hat{S}_i^\alpha-\langle\hat{S}_i^\alpha\rangle \langle\hat{S}_j^\alpha\rangle$ of the interactions across the spins inside and outside the cluster gives the effective one-body fields
\begin{eqnarray}
{\bm h}_i^{\rm MF}=\sum_{j\in \bar{C}}\left(J_{ij}\langle \hat{S}_j^x\rangle, J_{ij}\langle \hat{S}_j^y\rangle ,J^z_{ij}\langle \hat{S}_j^z\rangle\right),
\label{mf}
\end{eqnarray}
where $J_{ij}=J$ and $J_{ij}^z=J_z$ for nearest-neighbor pairs {$\langle i,j\rangle$} and $J_{ij},J_{ij}^z=0$ otherwise, and $\bar{C}$ is the outside of cluster $C$.

In contrast to the ED analysis in Sec.~\ref{sec2} with the periodic boundary condition, the existence of the mean fields ${\bm h}_i^{\rm MF}$, whose values are self-consistently determined, approximates the effects of spontaneous symmetry breaking expected to take place in the thermodynamic limit. Therefore, one can describe the distinction between the 0-coplanar and $\pi$-coplanar phases at the level of finite-size calculations. The approximation error caused by the mean-field decoupling can be neglected in principle by performing the cluster-size scaling $N_C\rightarrow \infty$ in the CMF+S method. Of course, there is a practical limitation on the cluster size $N_C$ that can be handled with numerical diagonalization. Unfortunately, since the assumption of the mean fields also breaks the conservation of the total spins $\sum_i \hat{S}_i^z$, the maximum size of {tractable clusters} becomes rather smaller than that of the ED analysis with the periodic boundary condition.

\begin{figure}[tb]
\includegraphics[scale=0.5]{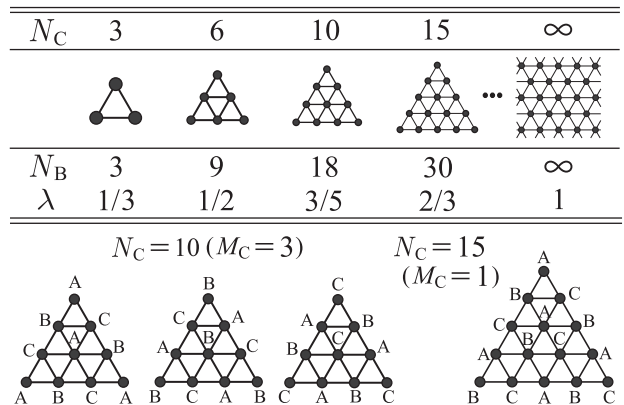}
\caption{\label{figclusters}
Series of the clusters used in the CMF+S analysis. The parameter $\lambda$ is defined by $\lambda\equiv N_B/(3N_C)$. The bottom illustrations show examples of the independent clusters that have to be considered under the three-sublattice ansatz. 
}
\end{figure}
In the following, we apply the CMF+S method to higher spins $S=1$ and $S=3/2$ to complement the study in Ref.~\cite{yamamoto-14} treating $S=1/2$. We calculate the transition points just below the saturation, $(J/J_z)_{{\rm c}1}$ between the 0-coplanar and $\pi$-coplanar phases and $(J/J_z)_{{\rm c}2}$ between the $\pi$-coplanar and umbrella phases, and compare the dependences on the spin value $S$ in the small-$S$ regime with those of the large-$S$ expansion given in Eq.~(\ref{largeS}).

In order to take an efficient scaling within the cluster-size limitation, we employ the series of triangular-shaped clusters shown in Fig.~\ref{figclusters}. The maximum size of the cluster in the practical calculations is $N_C=15$ for $S=1$ and $N_C=10$ for $S=3/2$. The magnetic orders such as 0-coplanar, $\pi$-coplanar, and umbrella are characterized by the sublattice magnetic moments $m^\alpha_\mu$ ($\alpha=x,y,z$) on sublattice $\mu={\rm A},{\rm B},{\rm C}$:
\begin{eqnarray}
m^\alpha_\mu=  \frac{1}{N_\mu}\sum_{n=1}^{M_C}\sum_{i_\mu\in C_n}\!{\rm Tr}\left( \hat{S}_{i_\mu}^{\alpha}e^{-\beta \hat{\mathcal{H}}_{C_n}}\right)\!\Big{/} {\rm Tr}( e^{-\beta \hat{\mathcal{H}}_{C_n}}), \label{selfconsistent}
\end{eqnarray}
where $i_\mu$ denotes site $i$ belonging to sublattice $\mu$, $M_C$ is the number of independent clusters determined by the matching between the cluster shape and the sublattice ansatz (see Fig.~\ref{figclusters}), $N_\mu$ is the number of total sites belonging to the sublattice $\mu$ in the $M_C$ clusters, and $\beta=1/T$ (we take $T\rightarrow 0$ to consider the ground-state properties). Substituting $m^\alpha_\mu$ into $\langle \hat{S}_{i_\mu}^{\alpha} \rangle$ in the effective field terms (${\bm h}_i^{\rm MF}$) of $\hat{\mathcal{H}}_{C_n}$, Eq.~(\ref{selfconsistent}) becomes a set of self-consistent equations for $m^\alpha_\mu$. Starting with a certain set of initial values for $m^\alpha_\mu$, we evaluate the right-hand side of Eq.~(\ref{selfconsistent}) in an iterative manner until the convergence of the values of $m^\alpha_\mu$ is reached. {Finally, we take the limit of the infinite cluster size for the results by using a scaling parameter $\lambda\equiv N_B/3N_C$~\cite{yamamoto-16,yamamoto-14,yamamoto-15,yamamoto-12-2}. Note that a similar approach has been also applied to study an $S=1$ bilinear-biquadratic model with single-ion anisotropy (at zero magnetic field)~\cite{moreno-cardoner-14}}.

\subsection{\label{}The CMF+S result for the transition points just below the saturation} 
Solving numerically Eq.~(\ref{selfconsistent}), we obtain the ground-state phase diagram of the triangular XXZ model for $S=1$ and $S=3/2$ in the strong-field regime, which has the same {topology} as in the case of $S=1/2$~\cite{yamamoto-14} [shown in Fig.~\ref{fig1}(b)]. {Both} phase transitions between the 0-coplanar and $\pi$-coplanar phases and between the $\pi$-coplanar and umbrella phases are of first order. At the transition points, the energies of each pair of the two phases are equal. We calculate the values of $J/J_z$ at the transitions with the help of the Maxwell construction in the plane of $J/J_z$ versus $\chi\equiv -\sum_{\langle i,j\rangle}\langle \hat{S}_i^x\hat{S}_j^x+\hat{S}_i^y\hat{S}_j^y \rangle/M$, {namely} the nearest-neighbor transverse correlation (see Ref.~\cite{yamamoto-14} for more {details}).

\begin{figure}[t]
\includegraphics[scale=0.6]{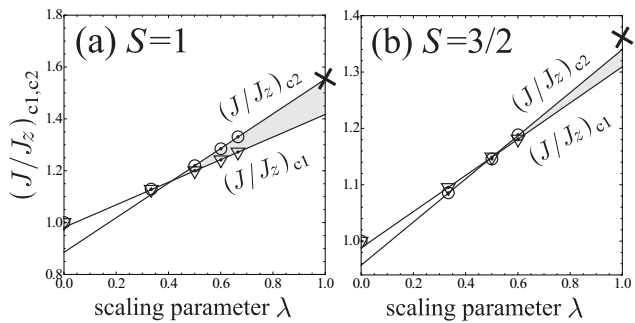}
\caption{\label{critical}
Cluster-size scaling of the CMF+S method for the phase boundaries just below the saturation field for (a) $S=1$ and (b) $S=3/2$. The transition points $(J/J_z)_{{c}1}$ and $(J/J_z)_{{c}2}$ correspond to the transitions between the 0-coplanar and $\pi$-coplanar phases and between the $\pi$-coplanar and umbrella phases, respectively. The symbols ($\times$) mark the semianalytical values {$(J/J_z)_{{c}2^*}$} from the {dilute Bose-gas} expansion~\cite{giacomo-16}}
\end{figure}
In Fig.~\ref{critical}, we show the cluster-size scaling of the phase transition points $(J/J_z)_{{\rm c}1}$ and $(J/J_z)_{{\rm c}2}$ just below the saturation field for $S=1$ and $S=3/2$. The scalings are performed with the parameter $\lambda\equiv N_B/3N_C$, which takes {values} of zero for $N_C=1$ and 1 for $N_C\rightarrow \infty$. Note that for $S=3/2$, the $\pi$-coplanar state is not energetically favorable against the $0$-coplanar or umbrella state for any $J/J_z$ when the cluster size is small ($N_C=3,6$). 
However, the $\pi$-coplanar phase is still found as a stationary solution of Eq.~(\ref{selfconsistent}). Therefore, in Fig.~\ref{critical} we plot the equal-energy points between the 0-coplanar and $\pi$-coplanar solutions and between the $\pi$-coplanar and umbrella solutions also for those small clusters in order to obtain a proper scaling series for $(J/J_z)_{{\rm c}1}$ and $(J/J_z)_{{\rm c}2}$.

\begin{table}[tb]
\begin{tabular}{cccc}
\hline
\hline
~~$S$~~& ~~$1/2$ ~~&~~$1$~~&~~$3/2$~~ \\ 
\hline 
CMF+S:~$(J/J_z)_{{\rm c}1}$~~ & ~~$1.588$~\cite{yamamoto-14} &~~$1.417$~~&~~$1.309$~~ \\
\hline 
CMF+S:~$(J/J_z)_{{\rm c}2}$~~ & ~~$2.220$~\cite{yamamoto-14} &~~$1.553$~~&~~$1.340$~~ \\
\hline 
{{dilute Bose-gas expansion}}:~~ & ~~$2.218$~~~~ &~~$1.554$~~&~~$1.361$~~ \\
~~{$(J/J_z)_{{\rm c}2^*}$}~\cite{giacomo-16}~~ & ~~~~~~ &~~~~&~~~~ \\
\hline 
\hline \\
\end{tabular}
\caption{\label{points}The CMF+S data for $(J/J_z)_{{\rm c}1}$ and $(J/J_z)_{{\rm c}2}$ obtained by the scaling with cluster sizes $N_C\leq 21$ for $S=1/2$~\cite{yamamoto-14}, $N_C\leq 15$ for $S=1$, and $N_C\leq 10$ for $S=3/2$. The semianalytical values for the coplanar-umbrella transition point {$(J/J_z)_{{\rm c}2^*}$} just below the saturation field~\cite{giacomo-16} are also listed for comparison. }
\end{table}
The fittings with a {linear} function are performed for the data points of the three largest clusters ($N_C=6,10,15$ for $S=1$ and $N_C=3,6,10$ for $S=3/2$). To see the accuracy, we also mark the semianalytical results of the arbitrary-$S$ {dilute Bose-gas} expansion for the coplanar-umbrella transition~\cite{giacomo-16}, which show good agreement with the extrapolated value of the $\pi$-coplanar-umbrella transition $(J/J_z)_{{\rm c}2}$ for $S=1$. The agreement is worse {(but still within 2 \%)} for $S=3/2$. This can be attributed to the fact that the scaling series includes the data of {the $N_C=3$ cluster, which is evidently too small}. Indeed, the fitting with a {linear} function does not seem {completely satisfactory} in this case.

For the 0-$\pi$ transition point $(J/J_z)_{{\rm c}1}$, no analytical value has been found in the literature. However, one can see that the linear fitting for $(J/J_z)_{{\rm c}1}$ is better than that for $(J/J_z)_{{\rm c}2}$. Even for $S=3/2$ with the data including $N_C=3$, the linear fitting is fine for $(J/J_z)_{{\rm c}1}$. One can clearly see that the difference $(J/J_z)_{{\rm c}2}-(J/J_z)_{{\rm c}1}$, i.e., the range where the $\pi$-coplanar phase appears, increases with $N_C\rightarrow \infty$ for both $S=1$ and $S=3/2$ as in the case of $S=1/2$~\cite{yamamoto-14}.

The CMF+S results for $(J/J_z)_{{\rm c}1}$ and $(J/J_z)_{{\rm c}2}$ in the small-$S$ regime are summarized in Table.~\ref{points} {together with the arbitrary-$S$ dilute Bose-gas results}~\cite{giacomo-16}. 
Although the {latter} has not addressed the distinction between {the 0-coplanar and $\pi$-coplanar states} for a {technical} reason (as was explained in Sec.~\ref{sec1}), the semi-analytical values for the transition point $(J/J_z)_{{\rm c}2^*}$ between the ({unspecified}) coplanar and umbrella phases are quantitatively reproduced as $(J/J_z)_{{\rm c}2}$ in our CMF+S method especially for $S=1/2$ and $S=1$, in which the calculations with relatively large clusters $N_C\geq 15$ have been done.

\subsection{\label{}Comparison with the large-$S$ approximation} 
\begin{figure}[t]
\includegraphics[scale=0.8]{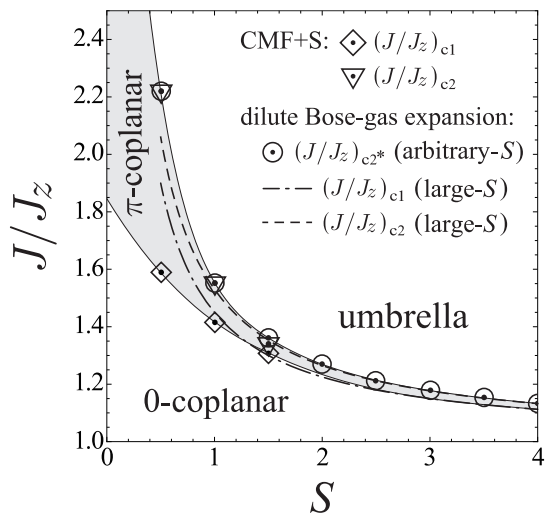}
\caption{\label{PhaseDiagram}
Phase boundaries among the two coplanar and umbrella states just below the saturation field. We show the current CMF+S results for $S=1$ and $S=3/2$ together with the previous $S=1/2$ values~\cite{yamamoto-14} for $(J/J_z)_{{\rm c}1}$ (inverted triangles) and $(J/J_z)_{{\rm c}2}$ (diamonds). {The {dilute Bose-gas} results for arbitrary $S$ (circles)~\cite{giacomo-16} and within the large-$S$ approximation (dashed and dashed-dotted curves)~\cite{starykh-14} are also plotted for comparison. Note that the arbitrary-$S$ {dilute Bose-gas} analysis provides a quantitatively precise value for the coplanar-umbrella transition point $(J/J_z)_{{\rm c}2^*}$ just below the saturation, but which coplanar state (0-coplanar or $\pi$-coplanar) appears for $J/J_z\leq (J/J_z)_{{\rm c}2^*}$ is unspecified~\cite{nikuni-95,giacomo-16}.} The boundary lines are just a guide for the eye. }
\end{figure}
Taking the current CMF+S results and the previous studies~\cite{yamamoto-14,giacomo-16,starykh-14} into account, we discuss the crossover from the nearly-classical, large-$S$ regime to the highly-quantum, small-$S$ regime of the magnetic phases in the TLAFs near saturation. In Fig.~\ref{PhaseDiagram}, we show the transition points $(J/J_z)_{{\rm c}1}$ and $(J/J_z)_{{\rm c}2}$ just below the saturation field as a function of spin $S$. One can see that the transition points for small $S\leq 3/2$, obtained by the CMF+S method, are smoothly connected to the large-$S$ values [Eq.~(\ref{largeS})] for $S\gtrsim 2$. It can be also expected that the $\pi$-coplanar phase is present as the ground state for any finite value of $S$, and asymptotically vanishes in the classical limit of $S\rightarrow \infty$.

Of particular interest is that the $J/J_z$ range where the $\pi$-coplanar phase is stabilized, i.e., $(J/J_z)_{{\rm c}2}-(J/J_z)_{{\rm c}1}$, monotonically increases as $S$ decreases. The most quantum case of $S=1/2$ has the greatest chance to observe the quantum stabilization of the $\pi$-coplanar phase in experiments on TLAF materials.

\section{\label{sec4}Conclusions} 
We have studied the ground-state magnetic phases as a function of anisotropy $J/J_z$ in the triangular-lattice XXZ model near saturation for small spins $S\leq 3/2$. In the former part, we {reconsidered} the previous ED analysis of Sellmann $et$ $al$. for $S=1/2$ in the three-magnon ($n=3$) sector of {the} Hilbert space~\cite{sellmann-15} by taking into consideration much larger-size clusters ($N\leq 1296$) and several higher eigenvalues. Moreover, we identified the associated eigenstates from the space-group symmetry and by calculating the overlaps with the coherent states corresponding to the candidate magnetic orders, i.e., the 0-coplanar, $\pi$-coplanar and umbrella states. For the system sizes $N=36,81,144,225$, the model exhibits three parameter $J/J_z$ ranges that have different lowest eigenstates. However, the intermediate one of the three ranges shrinks and vanishes {as $N$ is further increased}, or does not exist from the beginning in the case of another cluster series with $N=27,48,\cdots$. The authors in Ref.~\cite{sellmann-15} interpreted this result as indicating the nonexistence of the $\pi$-coplanar phase in the ground state, as opposed to our previous suggestion~\cite{yamamoto-14}. 

The current ED study with proper identification of the low-lying eigenstates offered a counterargument against Ref.~\cite{sellmann-15}. From the identification analysis, we demonstrated that the spurious phase which is present in finite-size calculations and disappears in the thermodynamic limit is actually a chirally symmetric combination of finite-size umbrella states. Furthermore, the lowest eigenstate in the coplanar region {$0<J/J_z\lesssim 2.218$} is always doubly degenerate{, and does not identify what coplanar state (more specifically, 0-coplanar or $\pi$-coplanar state) appears in the thermodynamic limit. Although the higher (but low-lying) eigenstates are expected to lift the degeneracy in such a case, it is also fundamentally impossible to argue that here since the collapse of the low-lying eigenstates into the symmetry-broken ground state in the thermodynamic limit $N\rightarrow\infty$ is much slower than the merging of the 0-coplanar and $\pi$-coplanar states into the magnetically saturated state as $n/N\rightarrow 0$. }

In the latter part of the paper, we applied the CMF+S method to the cases of $S=1$ and $S=3/2$ to complement the previous study~\cite{yamamoto-14}, in which the CMF+S has given a distinction between {the 0-coplanar and $\pi$-coplanar states} for $S=1/2$. Also for $S=1$ and $S=3/2$, we found the $\pi$-coplanar phase as the ground state for strong magnetic fields in a finite range of $J/J_z$ as well as the 0-coplanar and umbrella phases. In addition, we showed the crossover from the small-$S$ regime to the large-$S$ regime of the transition points just below the saturation field between the 0-coplanar and $\pi$-coplanar phases and between the $\pi$-coplanar and umbrella phases. It was predicted that the $\pi$-coplanar phase occupies the largest range of $J/J_z$ in the most quantum case of $S=1/2$, and asymptotically disappears as increasing $S$ towards the classical limit of $S\rightarrow \infty$.

In order to access the $\pi$-coplanar phase in real TLAF materials, it is required that the material has a relatively large easy-plane anisotropy, e.g., $J/J_z\approx 1.6$-$2.2$ for $S=1/2$, $J/J_z\approx 1.4$-$1.6$ for $S=1$, and $J/J_z\approx 1.3$-$1.4$ for $S=3/2$. A possible option is a family of Co-based compounds~\cite{shirata-12,zhou-12,lee-14,yokota-14,rawl-17}, which can possess an effective XXZ anisotropy due to the strong spin-orbit coupling. {For example, the latest estimation of the anisotropy parameter $J/J_z$ in Ba$_3$CoSb$_2$O$_9$ ($S=1/2$) ranges from $J/J_z\approx 1.18$ to $1.3$~\cite{sera-16,yamamoto-14,kamiya-17}.} Another way is to place TLAF materials under static pressure~\cite{ruegg-08,kurita-16}, which could tune the system parameters including the anisotropy. 

This work was supported by KAKENHI from Japan Society for the Promotion of Science: Grants No.~26800200 (D.Y.), No.~25800221 (H.U.), No.~25220711 (I.D.), and No.~16K05425 (T.M.) {and partially by CREST, JST Grant No. JPMJCR1673.} {The numerical computation was partially carried out on the Yukawa Institute Computer Facility, Kyoto University.}


\begin{thebibliography}{99}
\bibitem{note}
It should be noted that in Ref.~\cite{giacomo-16} there is an inaccuracy in the comparison with the $1/S$ approximation; this is due to a naive inversion of a quantity which is valid up to order $1/S$ without subsequently truncating the Taylor series, as is performed in {Eq.~(\ref{largeS})} instead. The appropriate truncation shows that the $1/S$ approximation is actually closer to the exact result, as apparent in Fig.~\ref{PhaseDiagram}.
\end{thebibliography}
\end{document}